\documentclass[prb,twocolumn]{revtex4}
\usepackage{graphicx}
\usepackage{dcolumn}
\usepackage{bm}
\usepackage{amsmath}
\usepackage{amsfonts}

\newcommand{\bra}[1]{\langle #1|}
\newcommand{\ket}[1]{|#1\rangle}
\newcommand{\braket}[2]{\langle #1|#2\rangle}

\begin{document}
\title{Quantum theory of electron tunneling  into intersubband cavity polariton states}

\author{Simone \surname{De Liberato}$^{1,2}$}
\author{Cristiano Ciuti$^{1}$}
\affiliation{$^1$Laboratoire Mat\'eriaux et Ph\'enom\`enes Quantiques, Universit\'e Paris
Diderot-Paris 7 et CNRS, UMR 7162, 75013 Paris, France} \affiliation{$^2$Laboratoire Pierre Aigrain,
\'Ecole Normale Sup\'erieure and CNRS, 24 rue Lhomond, 75005
Paris, France}

\begin{abstract}
Through a non-perturbative quantum theory,  we investigate how the
quasi-electron excitations of a two-dimensional electron gas are
modified by strong coupling to the vacuum field of a microcavity.
We show that the electronic dressed
states originate from a Fano-like coupling between the bare
electron states and the continuum of intersubband cavity polariton
excitations. In particular, we calculate the electron spectral
function modified by light-matter interactions and its impact on
the electronic injection of intersubband cavity polaritons. The
domain of validity of the present theoretical results is
critically discussed. We show that resonant electron tunneling
from a narrow-band injector can selectively excite superradiant
states and produce efficient intersubband polariton
electroluminescence.
\end{abstract}
\maketitle
\section{Introduction}
Cavity quantum electrodynamics in the strong coupling regime is presently the subject of many fascinating investigations in several interesting systems, including ultracold atoms\cite{Colombe},
Cooper pair quantum boxes \cite{Wallraff} and semiconductor nanostructures\cite{Hennessy}. In the strong coupling regime, the
eigenstates of a cavity system are a coherent mixing of photonic
and electronic excitations. This occurs when the light-matter
interaction, quantified by the so-called vacuum Rabi frequency, is
dominant with respect to loss mechanisms for the cavity photon
field and for the electronic transition.

Recently, the strong coupling regime has been demonstrated also between a planar microcavity mode and an intersubband transition in a doped semiconductor quantum well. The normal modes of such a system are called intersubband cavity polaritons \cite{Dini_PRL,Aji_APL,Aji_2006,SST,Ciuti_PRA,Ciuti_vacuum,Pere,Luca_APL,UltraEx}. The
active electronic transition is between two conduction subbands, where a dense two-dimensional electron gas populates
the lowest one. Large vacuum Rabi frequencies can be achieved thanks to the giant collective dipole
associated to the dense electron gas and even an unusual
ultra-strong coupling regime can be
reached\cite{Ciuti_PRA,Ciuti_vacuum, Simone_PRL}.

Electroluminescence experiments in microcavity-embedded quantum
cascade devices\cite{Luca_PRL,EL_model} have recently demonstrated
that it is possible to obtain intersubband cavity polariton
emission after resonant electrical excitation even at room
temperature (and even a lasing mechanism has been proposed
\cite{Scattering}). A fundamental question to address is how the
strong interaction with the microcavity vacuum field modifies the
quasi-electron states in the quantum well and how the electron
tunneling is affected. In this paper, we present a quantum theory
to investigate such fundamental problems. We show that the
electronic eigenstates originate from a Fano-like coupling between
the bare injected electron and the continuum of cavity polariton
modes. Our theory demonstrates that resonant electron tunnelling
from a narrow-band injector contact can selectively excite
polaritonic states leading to ultraefficient polariton
electroluminescence.

The paper is organized as follows: in Sec. II, we introduce the
general formalism, by presenting the quantum Hamiltonian,  the
approximations behind it and by introducing the concept of the
electron spectral function in the non-interacting case. In Sec.
III we present the calculations leading to the spectrum of the
Hamiltonian and to the determination of the electron spectral
function in the light-matter interacting case. In Sec. IV we
calculate the tunneling coupling and the radiative lifetime of the
excited states and present the resulting electroluminescence
spectra. Conclusions are drawn in Sec V.

\section{General formalism}

To describe the system to investigate, we consider
the following quantum Hamiltonian:
\begin{eqnarray}
\label{H} H&=&\sum_{\mathbf{k}}\hbar
\omega_{1}(k)c_{1,\mathbf{k}}^{\dagger}c_{1,\mathbf{k}}+
\sum_{\mathbf{k}}\hbar\omega_{2}(k)c_{2,\mathbf{k}}^{\dagger}c_{2,\mathbf{k}}+
\sum_{\mathbf{q}}\hbar\omega_{c}(q) a_{\mathbf{q}}^{\dagger}a_{\mathbf{q}} \nonumber \\
&&+\sum_{\mathbf{k,q}}\hbar\chi(q)
a_{\mathbf{q}}c_{1,\mathbf{k}}c_{2,\mathbf{k+q}}^{\dagger}
+\sum_{\mathbf{k,q}}\hbar\chi^{*}(q)
a_{\mathbf{q}}^{\dagger}c_{2,\mathbf{k+q}}c_{1,\mathbf{k}}^{\dagger}~,
\end{eqnarray}
where $\hbar \omega_{1}(k) = \frac{\hbar^2 \mathbf{k}^2}{2m^\star}$
and
 $\hbar \omega_{2}(k) = \hbar \omega_{12} + \frac{\hbar^2 \mathbf{k^2}}{2m^\star}$ are the energy dispersions of the two quantum well conduction subbands as a function of the in-plane wavevector $\mathbf{k}$, $m^\star$ being the effective mass.
 The corresponding electron creation operators are $c_{1,\mathbf{k}}^{\dagger}$ and $c_{2,\mathbf{k}}^{\dagger}$.
 $\omega_{c}(q)$ is the frequency dispersion of the cavity photonic mode and $a_{\mathbf{q}}^{\dagger}$
 is the corresponding photon  creation operator.
 Due to the selection rules of intersubband transitions,
 we omit the photon polarization, which is assumed to be Transverse Magnetic (TM).
 Being all the interactions spin-conserving, we can omit the electron spin.
 For simplicity, we consider only a photonic branch, which is quasi-resonant with the intersubband transition, while other cavity modes are supposed to be off-resonance.
The interaction between the cavity photon field and the two
electronic conduction subbands is quantified by the coupling constant
\begin{equation}
\chi(q) = \sqrt{\frac{\omega_{12}^2 d_{12}^2} {\hbar
\epsilon_0 \epsilon_r L_{cav} S \omega_c(q)}
\frac{q^2}{(\pi/L_{cav})^2 + q^2 }}~,
\end{equation}
where $c$ is the light speed, $d_{12}$ is the intersubband
transition dipole , $\epsilon_r$ is the cavity dielectric constant
and $S$ is the sample area. Here, we have considered the simple
case of a $\lambda/2$-cavity of lenght $L_{cav}$, where
$\pi/L_{cav}$ is the cavity photon quantized vector along the
growth direction. The geometrical factor
$\frac{q^2}{(\pi/L_{cav})^2 + q^2 }$ is due to the TM-polarization
nature of the intersubband transition. Here, for simplicity, we
are considering the case of just a single quantum well coupled to
the cavity quantum field. Notice that in the Hamiltonian in Eq.
(\ref{H}) the anti-resonant terms of the light-matter interaction
have not been included. Therefore, here we can describe the strong
coupling  for the two subband system, but not all the peculiar
features of the ultrastrong coupling regime
\cite{Ciuti_vacuum,Ciuti_PRA,Simone_PRL,UltraEx}.
 
 If one is interested in describing the dynamics of the microcavity
under optical excitation, it is possible to use an effective
Hopfield Hamiltonian with bosonic operators associated to the
intersubband polaritons, which are the elementary optical
excitations of the system. If, instead, one is interested in
studying microscopically how the electronic injection into to such
a microcavity system is modified by non-perturbative light-matter
excitation, it is necessary to work with the full fermionic
Hamiltonian in Eq. (\ref{H}). In fact, electrical excitation
occurs through injection of (fermionic) carriers: the dynamics
must include the (bosonic) optical excitations and the electronic
(fermionic) excitations at the same level. As well known in the
theory of quantum transport\cite{Datta}, if we wish to study the
tunneling injection of one electron at low temperature, we have to
determine the electron spectral function, defined as:
\begin{equation}
\label{spectraldensity} A^+_j(\mathbf{k},\omega)=\sum_{\zeta}
\lvert \bra{\zeta} c^{\dagger}_{j, \mathbf{k}}\ket{F_N} \lvert
^2\delta(\omega -\omega_{\zeta}) ~,
\end{equation}
where $\ket{F_N}$ is the N-electron Fermi sea ground state times
the vacuum state for the cavity photon field and  $j=1,2$ is the
conduction subband index. The index $\zeta$ labels the excited
(N+1)-electron eigenstates and $\hbar\omega_{\zeta}$ are the
corresponding eigenenergies. Note that even if in the present
paper we consider the case of zero temperature, the theory can be
applied as far as $K_B T$ is much smaller than the Fermi energy of
the two-dimensional electron gas.

As apparent from Eq. (\ref{spectraldensity}), the electron
spectral function is the density of quasi-electron states,
weighted by the overlap with the bare electron state
$c^{\dagger}_{j, \mathbf{k}}\ket{F_N}$. In other words, it is the
many-body equivalent of the single electron density of states.
This is the key quantity affecting the electron tunneling and can
be non-trivially modified by  interactions like in the case of
superconductors. For a non-interacting electron gas,
$c^{\dagger}_{1, \mathbf{k}}\ket{F_N}$ and $c^{\dagger}_{2,
\mathbf{k}}\ket{F_N}$ are eigenstates of the Hamiltonian and thus
all the other eigenstates are orthogonal to them. Therefore the
non-interacting spectral functions are

\begin{eqnarray}
\label{A1}
A^{+}_1(\mathbf{k},\omega)&=& \delta(\omega -\omega_1(k))\theta(k-k_F),\\
&&\nonumber\\
 \label{A2}
A^{+}_2(\mathbf{k},\omega)&=& \delta(\omega -\omega_2(k)), 
\end{eqnarray}

where $k_F$ is the Fermi wavevector. $\theta(x)$ is the Heaviside
function and its presence is due to Pauli blocking:
$c^{\dagger}_{1, \mathbf{k}}\ket{F_N}=0$ for $k<k_F$.

\section{Spectral function with light-matter interactions}
As seen in the previous section, in the non-interacting case, the
electron spectral function is just a Dirac delta. Physically, this
means that an electron with wavevector $\mathbf{k}$ can be
injected in the subband $j=1,2$ only with an energy equal to the
bare electron energy $\hbar\omega_j(k)$ and that such excitation
has an infinite lifetime. By contrast, interactions can profoundly
modify the nature of electron excitations and therefore produce
qualitative and quantitative changes of the electron spectral
functions. In the case of a weakly interacting electron gas, the
spectral function has a broadened "quasi-electron" peak: the
spectral broadening is due to the finite lifetime of the
electronic excitation. In the case of a strongly interacting
electron gas (like in the case of superconductors) the electron
spectral function can be qualitatively different from the
non-interacting gas. Here, we are interested in how the nature of
the quasi-electron excitations is modified by the strong coupling
to the vacuum field of a microcavity. In particular, we assume
that the light-matter interaction is the strongest one. We will
provide here a nonperturbative theory to determine the dressed
electronic excitations in such a strong coupling limit and their
corresponding spectral function. All other residual interactions
will be treated as perturbations. The consistency and limit of
validity of such a scheme will be discussed in the next section,
where the theoretical results are applied.

 In the interacting case, it is easy to verify that
$c^{\dagger}_{1, \mathbf{k}}\ket{F_N}$  is still an eigenvector
of the Hamiltonian in Eq. (\ref{H}) and thus the first subband
spectral function $A_1^+(\mathbf{k},\omega)$ is still given by Eq.
(\ref{A1}). Instead for the electrons in the second subband we
have to distinguish between two cases: $\mathbf{k}$ well inside or
outside the Fermi sea. In the first case, an electron in the
second subband can not emit a photon because all the final states
in the first subband are occupied (Pauli blocking), hence the
spectral function will be given by the unperturbed one (Eq.
(\ref{A2})). Well outside the Fermi sea,  an injected electron can
emit and the spectral function will be modified by the
light-matter interaction. A smooth transition between the two
cases will take place for  $\lvert k-k_F \lvert $ of the order of
the resonant cavity photon wave-vector $q_{res}$, where
$\omega_{c}(q_{res}) = \omega_{12}$. Being the ratio $q_{res}/k_F$
typically  very small, of the order of $10^{-2}$ (see Ref.
[\onlinecite{Aji_excited}]), we can safely consider an abrupt
transition at the Fermi edge.

\begin{figure}[t!]
\begin{center}
\includegraphics[width=9cm]{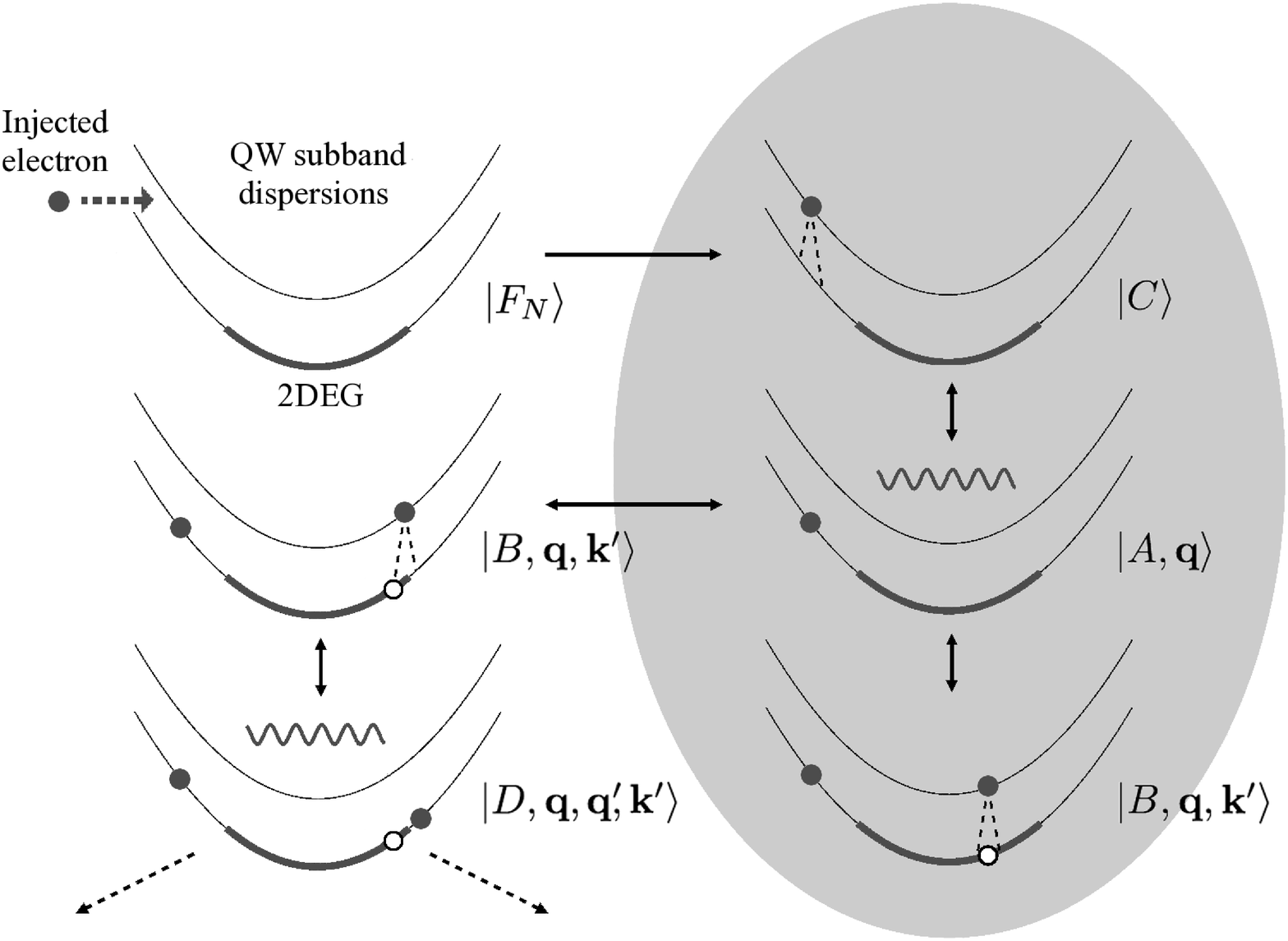}
\caption{\label{fig1} Sketch of the dynamical coupling between
quantum states in a microcavity-embedded quantum well (QW)
containing a two-dimensional electron gas (2DEG). In the ground
state, the first subband is doped with a dense 2DEG (bold lines at
the bottom of the dispersions).  Black dots represent bare
electrons, while white dots denote holes in the 2DEG. The dashed
cones depicts the possible final states for an electron
radiatively relaxing from the second to the first subband by
emission of a cavity photon. The ground state with $N$ electrons
is the standard Fermi sea $\ket{F_N}$. The injection (e.g.,
through electron tunneling) of an additional electron in the
second subband  creates the state
$\ket{C}=c^{\dagger}_{2,\mathbf{k}}\ket{F_N}$, which, in presence
of light-matter interaction, is not an eigenstate. Spontaneous
emission of a cavity photon couples the $\ket{C}$ state to the
states $\ket{A,\mathbf{q}}=a^{\dagger}_{\mathbf{q}}
c^{\dagger}_{1, \mathbf{k-q}}\ket{F_N}$. Reabsorption of the
emitted cavity photon can couple back to the  $\ket{C}$ state or
to the states $\ket{B,\mathbf{q,k'}}=c^{\dagger}_{2,
\mathbf{k'+q}}c_{1, \mathbf{k'}}c^{\dagger}_{1,
\mathbf{k-q}}\ket{F_N}$. Spontaneous emission couples the
$\ket{B}$ states back to $\ket{A}$ states or to states of the form
$\ket{D,\mathbf{q,q',k'}}=a^{\dagger}_{\mathbf{q'}}c_{1,
\mathbf{k'+q-q'}}^{\dagger}c_{1, \mathbf{k'}}c^{\dagger}_{1,
\mathbf{k-q}}\ket{F_N}$. Being the relevant cavity photon
wavevectors very small compared to the Fermi wavevector,
spontanous emission can occur only on narrow emission cone in
momentum space. Due to the small probability of photon absorption
by electrons on the border the Fermi sea, we can neglect $\ket{D}$
states and assume that the system always jumps from $\ket{B}$
states to $\ket{A}$ states. Thus the relevant dynamics takes place
only between the states in the shaded region. We can thus neglect
the other marginal states while diagonalizing the light-matter
Hamiltonian.}
\end{center}
\end{figure}

For evaluating $A_2^+(\mathbf{k},\omega)$ for $k>k_F$ we
need to find all the (N+1)-electron eigenstates that have a
nonzero overlap with $c^{\dagger}_{2, \mathbf{k}}\ket{F_N}$. In
order to do this we notice that the Hamiltonian in Eq. (\ref{H})
commutes with the number of total fermions

\[ \hat{N}_F = \sum_{j=1,2} \sum_{\mathbf{k}}
c_{j,\mathbf{k}}^{\dagger}c_{j,\mathbf{k}},\]

the total in plane wave-vector operator

\[\hat{\mathbf{K}} =  \sum_{j=1,2}
\sum_{\mathbf{k}} \mathbf{k}
~c_{j,\mathbf{k}}^{\dagger}c_{j,\mathbf{k}}+\sum_{\mathbf{q}}\mathbf{q}a^{\dagger}_{\mathbf{q}}a_{\mathbf{q}},\]

and the excitation number operator

\[\hat{Q}=\sum_{\mathbf{k}}
c_{2,\mathbf{k}}^{\dagger}c_{2,\mathbf{k}}+\sum_{\mathbf{q}}a_{\mathbf{q}}^{\dagger}a_{\mathbf{q}}.\]
 Hence the
eigenstates $\ket{\zeta}$ of $H$ can be also  labeled by the corresponding eigenvalues
$N_{\zeta}, \mathbf{K_{\zeta}}$ and $Q_{\zeta}$. We will thus
identify an eigenstate of $H$ in the subspace
$(\hat{N_F}=N,\mathbf{\hat{K}=\mathbf{K}},\hat{Q}=Q)$ as
$\ket{N,\mathbf{K},Q,\zeta}$, where the index $\zeta$ now runs
over all the eigenstates of the subspace
The states obtained by applying electron creation or
destruction operators on the eigenstates $\ket{{N,\mathbf{K},Q,\zeta}}$ are still eigenstates of
$\hat{N}_F$, $\hat{\mathbf{K}}$ and $\hat{\mathbf{Q}}$. The state $c_{1,\mathbf{k}}^{\dagger} \ket{{N,\mathbf{K},Q,\zeta}}$ is in the subspace labeled by the quantum numbers $(N+1, \mathbf{K+k}, Q)$;
$c_{1,\mathbf{k}} \ket{{N,\mathbf{K},Q,\zeta}}$ in $(N-1, \mathbf{K-k}, Q)$;
$c_{2,\mathbf{k}}^{\dagger}\ket{{N,\mathbf{K},Q,\zeta}}$ in $(N+1, \mathbf{K+k}, Q+1)$;
$c_{2,\mathbf{k}}\ket{{N,\mathbf{K},Q,\zeta}}$ in $(N-1, \mathbf{K-k},Q-1)$.

Having $\ket{F_N}$ quantum numbers (N,{\bf 0},0) the state $c_{2,\mathbf{k}}^{\dagger} \ket{F}$ is thus in the subspace labeled by the quantum numbers $(N+1, \mathbf{k}, 1)$. We can limit ourselves to diagonalize $H$ in this subspace, which
 is spanned by vectors of the form: (i)
$c^{\dagger}_{2,\mathbf{k}_0} \prod_{j=1}^N
c^{\dagger}_{1,\mathbf{k}_j} \ket{0}$, where $\ket{0}$ is the
empty conduction band state and $\sum_{j=1}^N \mathbf{k}_j
 = \mathbf{k}- \mathbf{k}_0$; (ii) $a^{\dagger}_{\mathbf{q}_0}
 \prod_{j=1}^{N+1} c^{\dagger}_{1,\mathbf{k}_j} \ket{0}$ with
 $\sum_{j=1}^{N+1} \mathbf{k}_j
 = \mathbf{k}- \mathbf{q}_0$. For a large number of electrons, the exact diagonalization of the Hamiltonian in this subspace
 is an unmanageable task. Here, we show
 that by a judicious approximation, we can considerably simplify
 the diagonalization problem, keeping the relevant
 non-perturbative physics. Namely, we claim that the elements of the ($N+1,\mathbf{k}, 1$) subspace can be well approximated by vectors of the form
\begin{widetext}
\begin{equation}
\label{general_vector} \ket{N+1,\mathbf{k}, 1,\zeta}= \left \{
\mu_{\zeta}~ c^{\dagger}_{2, \mathbf{k}} +  \sum_{\mathbf{q}} \left
[ \alpha_{\zeta}(\mathbf{q})~ a^{\dagger}_{\mathbf{q}}
c^{\dagger}_{1, \mathbf{k-q}} +\sum_{\mathbf{\lvert k'
\lvert}<k_F}\beta_{\zeta}~(\mathbf{q , k'}) c^{\dagger}_{2,
\mathbf{k'+q}}c_{1, \mathbf{k'}}c^{\dagger}_{1, \mathbf{k-q}}
\right] \right \} \ket{F_N}~.
\end{equation}
\end{widetext}

\begin{figure}[t!]
\begin{center}
\includegraphics[width=9.3cm]{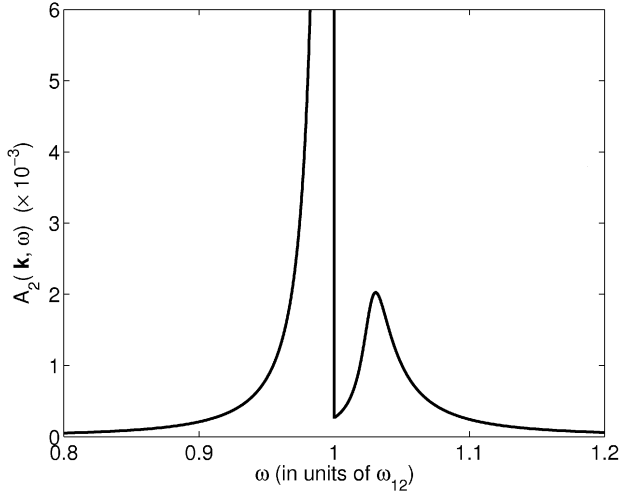}
\caption{\label{fig2}  Electron spectral function
$A^+_2(\mathbf{k},\omega)$ for the second subband , for all
$k>k_F$. The spectral function, defined in Eq. (\ref{spectraldensity}), is the density of quasi-electron states, weighted by the overlap with the bare electron state.The integral of the spectral function over the frequency $\omega$ is by construction equal to one. Coupling parameter: $\Omega_R(q_{res}) = \lvert \chi(q_{res})\lvert  \sqrt{N} =
0.1 \omega_{12}$.  }
\end{center}
\end{figure}
To understand the origin of our approximation, let us consider the
time evolution picture sketched in Fig. \ref{fig1}. Suppose that
initially the system is in its ground state $\ket{F_N}$. After
injection of one bare electron, the state of the system is
\[\ket{C}=c^{\dagger}_{2, \mathbf{k}}\ket{F_N}.\] If $\mathbf{k}$
is well inside the Fermi sphere, as we said before, it is Pauli blocked and can
not radiatively relax into the first
subband.  Instead, when $k > k_F$, the injected electron can
radiatively decay, emitting a photon and falling into the first
subband. After the first emission the state will have the form
\[\ket{A,\mathbf{q}}=a^{\dagger}_{\mathbf{q}} c^{\dagger}_{1,
\mathbf{k-q}}\ket{F_N}.\] If the cavity system is closed and only
the light-matter interaction is considered, the emitted photon
will be eventually reabsorbed. The system can evolve back to the state
$\ket{C}$ or into one vector of the form
\[\ket{B,\mathbf{q,k'}}=c^{\dagger}_{2, \mathbf{k'+q}}c_{1,
\mathbf{k'}}c^{\dagger}_{1, \mathbf{k-q}}\ket{F_N}.\] If
$\mathbf{k'}$ is well inside the Fermi sea, when the second
subband electron decays, the only avaiable final state
 in the first subband will be the one  with wavevector $\mathbf{k'}$,
that is the system will go back to state $\ket{A,\mathbf{q}}$. If
$\mathbf{k'}$ is on the border of the Fermi sea, on the contrary,
the system can evolve into a state of the form
$\ket{D,\mathbf{q,q',k'}}=a^{\dagger}_{\mathbf{q'}}c_{1,
\mathbf{k'+q-q'}}^{\dagger}c_{1,\mathbf{k'}}c^{\dagger}_{1,
\mathbf{k-q}}\ket{F_N}$. The probability of ending in
any of the $\ket{D,\mathbf{q,q',k'}}$ states is negligible. In
fact, the probability for $\mathbf{k'}$ to be near enough to the
border of the Fermi sea for allowing an emission to electronic
states with $k>k_F$ is proportional to the ratio $q_{res}/k_F\ll
1$. Hence, the diagonalization problem can be simplified and we
can thus look for vectors of the form shown in Eq.
(\ref{general_vector}).

\begin{widetext}

In such a subspace, spanned by $\{ \ket{C}, \ket{A,\mathbf{q}
},\ket{B,\mathbf{q,k'}},\ket{B,\mathbf{q,k''}},\dots,\ket{A,\mathbf{q'}
},\ket{B,\mathbf{q',k'}},\ket{B,\mathbf{q',k''}},\dots\}$, $H$ has the following
matrix representation:

\[
{\mathcal H}_{N+1,\mathbf{k}, 1}= \hbar \left(
\begin{array}{ccccc}
\omega_{2}(k)  & v(q)   & v(q') & v(q'') & \cdots  \\
 v(q)^T &  M(\mathbf{q}) & 0 &  0 & \cdots \\
 v(q')^T &  0 & M(\mathbf{q'}) & 0 & \cdots \\
 v(q'')^T &  0 & 0 & M(\mathbf{q''}) & \ddots \\
 \vdots & \vdots & \vdots & \ddots & \ddots\\
\end{array}
\right)
\]

where $M(\mathbf{q})$ is the Hamiltonian matrix block in the
subspace spanned by
$\{\ket{A,\mathbf{q}},\ket{B,\mathbf{q,k'}},\ket{B,\mathbf{q,k''}},\dots\}$,
that effectively describes the system in presence of one photon
with a well defined wavevector $\mathbf{q}$

\[
M(\mathbf{q})=\left(
\begin{array}{cccc}
\omega_c(q)+\omega_{1}(|\mathbf{k}-\mathbf{q}|)) & \chi^{*}(q)  & \chi^{*}(q) & \cdots \\
\chi(q)  & \omega_{2}(|\mathbf{k'}+\mathbf{q}|) - \omega_{1}(k')+ \omega_{1}(|\mathbf{k}-\mathbf{q}|)  &  0 & \cdots \\
\chi(q)  &  0 &  \omega_{2}(|\mathbf{k''}+\mathbf{q}|) - \omega_{1}(k'') + \omega_{1}(|\mathbf{k}-\mathbf{q}|) & \ddots \\
\vdots & \vdots & \ddots & \ddots \\
\end{array}
\right),
\]

\end{widetext}

where $v(q)=[\chi(q) ,0, 0, \dots ]$. Since the typical
wavevector $q$ of the resonantly coupled cavity photon mode is
much smaller than $k_F$, we can perform the standard approximation
$\omega_{2}(|\mathbf{k'}+\mathbf{q}|)-\omega_{1}(k') \simeq
\omega_{2}(k')-\omega_{1}(k') = \omega_{12}$. This way, we can
exactly diagonalize each of the $M(\mathbf{q})$ blocks. As
expected from the theory of optically excited polaritons
\cite{Ciuti_vacuum}, by diagonalizing the matrix $M(\mathbf{q})$
we find two bright electronic states (i.e., with a photonic mixing
component)
\begin{eqnarray}
\ket{\pm,\mathbf{q}}&=&\frac{(\omega_{\pm}(q)-\omega_{12})\ket{A,\mathbf{q}}
+\chi(q)\sum_{\mathbf{k}}\ket{B,\mathbf{q,k}}} {\sqrt{\left (
\omega_{\pm}(q)-\omega_{12} \right)^2+\lvert \chi(q)\lvert ^2N}}~,
\end{eqnarray}
with energies $\hbar \omega_{1}(k) + \hbar \omega_{\pm}(q)$, where
\begin{equation}
\omega_{\pm}(q)=\frac{\omega_c(q)+\omega_{12}}{2}\pm\sqrt{\left
(\frac{\omega_c(q)-\omega_{12}}{2} \right)^2+N\lvert
\chi(q)\lvert^2}~.
\end{equation}
Note that $\hbar \omega_{\pm}(q)$ are the energies of the two
branches of intersubband cavity polaritons \cite{Ciuti_vacuum}.

The other orthogonal states are dark (no photonic component), with
eigenvalues $\omega_{2}(k)=\omega_{1}(k) + \omega_{12}$ and
eigenvectors
\begin{eqnarray}
\ket{l,\mathbf{q}}&=&\frac{\sum_{\mathbf{k}}\beta_{l}(\mathbf{q , k})\ket{B,\mathbf{q,k}}}{\sqrt{N}}
\end{eqnarray}
where the $\beta_{l}(\mathbf{q , k})$ are such that $\sum_{\mathbf{k}} \beta_{l}(\mathbf{q , k})=0$ and
$\sum_{\mathbf{k}} \beta_{l}(\mathbf{q , k})\beta_{l'}^*(\mathbf{q
, k})=\delta_{l,l'}$.
\begin{figure}[t]
\begin{center}
\includegraphics[width=9.3cm]{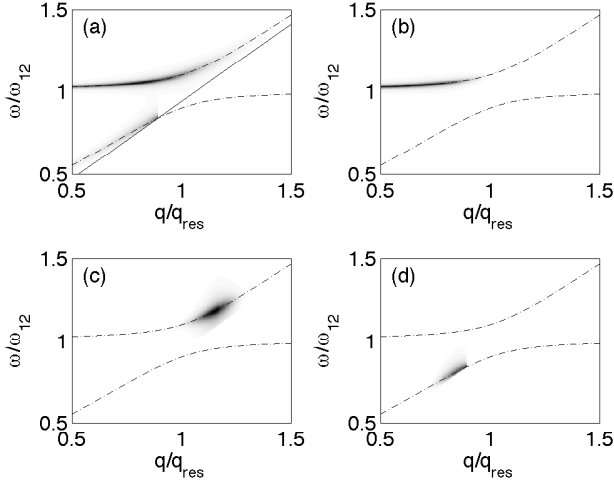}
\caption{\label{fig3}Extracavity electroluminescence spectra
$N_{ph}(\mathbf{q}, \omega)$. Panel (a): the case of a broadband
electrical injector (bandwidth equal to $\omega_{12}$, centered at
$\omega = \omega_{12}$). The other panels show the results for a
narrow-band injector (width $0.05\omega_{12}$) centered
respectively at $\omega = \omega_{12}$ (b), $1.2\omega_{12}$ (c) and
$0.8\omega_{12} (d)$. The non-radiative relaxation rate $1/\tau_{nr,\mathbf{k},\zeta}$
has been taken equal to $0.005 \omega_{12}$. In all panels, the
dashed-dotted lines are the frequency dispersions $\omega_{\pm}(q)$
of the two cavity polariton branches. In the first panel the solid
line represents the edge of the light cone \cite{Aji_excited}.}
\end{center}
\end{figure}

Since $\bra{l,\mathbf{q}}  H~ c_{2,\mathbf{k}}^{\dagger}
\ket{F_N}=0$,
the dark states $\ket{l,\mathbf{q}}$ are also eigenstates of the
matrix ${\mathcal H}_{N+1,\mathbf{k},1}$ and do
not contribute to the electron spectral function, because they
have zero overlap with the state $\ket{C}=c_{2,\mathbf{k}}^{\dagger}
\ket{F_N}=0$. In contrast, this is not the case for the bright eigenstates of
each block $M({\mathbf{q}})$, as we find:
\begin{equation}
\bra{\pm,\mathbf{q}}  H~ c_{2,\mathbf{k}}^{\dagger}
\ket{F_N}=\frac{\chi(q)^{*}(\omega_{\pm}(q)-\omega_{12})}
{\sqrt{(\omega_{\pm}(q)-\omega_{12})^2+\lvert \chi(q)\lvert^2N}}
=J_{\pm}^{*}(q)~.
\end{equation}
Therefore, the representation of $H$ in the subspace
$\{ \ket{C}, \ket{+,\mathbf{q}},
\ket{-,\mathbf{q}}, \ket{+,\mathbf{q'}}, \ket{-,\mathbf{q'}},
\dots \}$ reads

\begin{widetext}
\[
{\mathcal H'}_{N+1,\mathbf{k}, 1}=
  \hbar \left(
\begin{array}{cccccc}
\omega_{1}(k) + \omega_{12}                    & J_{+}^{*}   (q)          & J_{-}^{*}   (q)    & J_{+}^{*}   (q')        & J_{-}^{*}(q') & \cdots\\
J_{+}   (q)  & \omega_{1}(k) + \omega_+(q)          &  0                                             & 0                                              & 0 & \cdots \\
J_{-}   (q)  &  0                                               & \omega_{1}(k) + \omega_-(q)       & 0                                               & 0 & \cdots\\
J_{+}   (q')   &  0                                              &  0                                              & \omega_{1}(k) + \omega_+(q')   & 0 & \cdots   \\
 J_{-}   (q')  &  0                                               & 0                                              & 0  & \omega_{1}(k) +\omega_-(q') & \cdots \\
 \vdots & \vdots & \vdots & \vdots & \vdots & \ddots \\

\end{array}
\right).
\]
\end{widetext}
Hence, here we have demonstrated that the bare electron state
$c_{2,\mathbf{k}}^{\dagger} \ket{F_N}$ is coupled to the continuum
of the polariton modes with all the different wavevectors
$\textbf{q}$. Since the polariton frequencies $\omega_{\pm}$ and the
coupling $J_{\pm}$ depend only on the modulus of $\textbf{q}$, we
can further simplify the diagonalization problem by introducing
the 'annular' bright states
\begin{eqnarray}
\ket{\pm, q}=\frac{1}{\sqrt{Lq}}\sum_{\lvert \mathbf{q} \lvert =
q} \ket{\pm, \mathbf{q}}~,
\end{eqnarray}
where $L = \sqrt{S}$ and $2\pi/L$ is the linear density of modes in reciprocal space.
All annular states are coupled to
$\ket{C}$. Instead, all the orthogonal linear combinations of
$\ket{\pm, \mathbf{q}}$ (with $\lvert \mathbf{q} \lvert = q$) are
uncoupled and therefore do not contribute to the electron spectral
function. The matrix representation of $H$ in the subspace
$\{\ket{C}, \ket{+,q}, \ket{-,q},
\ket{+,q'}, \ket{-,q'}, \dots \}$ reads

\begin{widetext}
\begin{equation}
{\mathcal H''}_{N+1,\mathbf{k}, 1}= \hbar \left(
\begin{array}{cccccc}
 \omega_{1}(k) + \omega_{12}                 & J_{+}^{*}   (q)\sqrt{Lq }          & J_{-}^{*}   (q)\sqrt{L q }    & J_{+}^{*}   (q') \sqrt{L q' }          & J_{-}^{*}   (q') \sqrt{L q' } & \cdots \\
J_{+}   (q)\sqrt{L q } & \omega_{1}(k) + \omega_+(q)          &  0
& 0
& 0 & \cdots \\
J_{-}   (q) \sqrt{Lq } &  0                                               &   \omega_{1}(k) + \omega_-(q)       & 0                                               & 0 & \cdots \\
J_{+}   (q') \sqrt{L q' }  &  0                                              &  0                                              &  \omega_{1}(k) +\omega_+(q')          & 0 & \cdots \\
 J_{-}   (q') \sqrt{L q' }&  0                                               & 0                                              & 0                                                &
\omega_{1}(k) + \omega_-(q') & \ddots \\
\vdots & \vdots & \vdots & \vdots & \ddots & \ddots
\end{array}
\right) \label{finalMatrix}
\end{equation}
\end{widetext}

Hence, in the subspace ($N+1,\mathbf{k}, 1$), we have found that eigenstates of $H$ with a finite overlap with the bare electron have the form
\begin{equation}
\ket{N+1,\mathbf{k}, 1,\zeta} =
\mu_{\zeta}~ c^{\dagger}_{2\, \mathbf{k}} \ket{F_N}+  \sum_{q, \sigma=\pm}  \lambda_{\zeta,\sigma,q} \ket{\sigma,q}~.
\end{equation}
The coefficients $\mu_{\zeta}$ and $\lambda_{\zeta,\sigma,q} $ as well as the corresponding energy eigenvalues $\hbar\omega_{\zeta}$ can be calculated though a numerical diagonalization of the matrix in Eq. (\ref{finalMatrix}). In conclusion, the spectral function of the second subband reads
\begin{equation}
\label{spectral}
A^+_2(\mathbf{k},\omega)=\sum_{\zeta}
\lvert \mu_{\zeta} \lvert
^2\delta( \omega -\omega_{\zeta}) \theta (k - k_F) +
\delta( \omega - \omega_2(k)) \theta (k_F - k).
\end{equation}
In Fig. \ref{fig2}, we show numerical results using a vacuum Rabi
frequency $\Omega_R(q_{res}) = \lvert \chi(q_{res}) \rvert \sqrt{N} =
0.1\omega_{12}$. As it appears from Eq. (\ref{spectral}),  the
broadening of the spectral function is intrinsic, being associated
to the continuum spectrum of frequencies $\omega_{\zeta}$
corresponding to the dressed electronic states and given by the
eigenvalues of the infinite matrix in Eq. (\ref{finalMatrix}).
At each frequency
$\omega_{\zeta}$, the magnitude of the spectral function is given
by the spectral weight $\lvert \mu_{\zeta} \lvert^2$, depending on
the overlap between the dressed state $\ket{N+1,\mathbf{k},
1,\zeta}$ and the bare electron state $\ket{C}=c^{\dagger}_{2,
\mathbf{k}} \ket{F_N}$. As shown in Eq. (\ref{finalMatrix}), the
electronic eigenstates of the system are given by the Fano-like
coupling between the bare electron state and the continuum of
cavity polariton excitations. Indeed, the pronounced dip around $\omega = \omega_{12}$ in the
spectral function is a quantum interference feature, typical of a
Fano resonance\cite{Fano}.

 As we said before the sharp transition in Eq. \ref{spectral} between $k>k_F$ and $k<k_F$ is only
 a consequence of the approximations we made of neglecting the {\it border} of the Fermi sea and the effect of the temperature.
 In a real case both effects will tend to smooth the transition, the first on an energy scale of the order of $\frac{\hbar^2 k_F q_{res}}{m^*}$ and the second on an energy scale of $K_BT$.

\section{Tunneling coupling, losses and electroluminescence}

The states $\ket{N+1,\mathbf{k}, 1,\zeta}$ have been obtained by
diagonalizing the  Hamiltonian (\ref{H}), which takes into
account only the coupling between the two-subband electronic
system and the microcavity photon quantum field. If, as we have
assumed, the light-matter interaction is the strongest one, all
other residual couplings can be treated perturbatively. These
residual interactions include the coupling to the
extracavity fields, the interaction with contacts, phonon and
impurity scattering as well as Coulomb electron-electron
interactions\cite{Coulomb}.

The states $\ket{N+1,\mathbf{k}, 1,\zeta}$ can be excited for
example by resonant electron tunneling  from a bulk
injector or an injection miniband. If $V^{tc}_{\mathbf{k}}$ is the
tunneling coupling matrix element between the state $\ket{F}$ and
$c^{\dagger}_{2, \mathbf{k}}\ket{F}$ induced by the coupling with
the injector we have, using the Fermi golden rule, the following
injection rate:

\begin{eqnarray}
\label{injector} \Gamma_{inj}(\mathbf{k}, \zeta
)=\frac{2\pi}{\hbar} \lvert \mu_\zeta \lvert ^2 \lvert
V^{tc}_\mathbf{k}\lvert^2\rho_{inj}( \omega_\zeta)n_F(
\omega_\zeta ),
\end{eqnarray}

where $\rho_{inj}(\omega)$ is the density of electronic states
inside the contact and $n_f(\omega)$ its Fermi distribution. $\rho_{inj}(\omega)n_f(\omega)$ determines the spectral shape of the injector. $\mu_\zeta$ comes from Eq. (\ref{spectral}) and represents the
electron spectral weight.

It is worthwhile to notice that the formula in Eq. \ref{injector} is quite independent from the model of injector considered. All the relevant informations are contained in the coupling strenght $V^{tc}_{\mathbf{k}}$ and the spectral shape $\rho_{inj}(\omega)n_f(\omega)$.
Any form of scattering, including in-plane wavevector non-conserving interactions or non-resonant injection, will simply give a different (and possibly broadened) injector spectral shape.

The finite transmission of the cavity mirrors is responsible for a
finite lifetime for the cavity photons and consequently for the
dressed states $\ket{N+1,\mathbf{k}, 1,\zeta}$. By using the Fermi
golden rule and a quasi-mode coupling to the extracavity field, we
find that the radiative lifetime $\tau_{r,\mathbf{k}, \zeta}$  reads:
\begin{eqnarray*}
\label{taur} \frac{1}{\tau_{r,\mathbf{k},
\zeta}}&=&\frac{2\pi}{\hbar}\sum_{\mathbf{q}, q_z}
|\alpha_{\zeta}(\mathbf{q}) \lvert^2 \lvert  V^{qm}_{\mathbf{q},q_z}\lvert^2
\delta(\hbar \omega_{\zeta}- \hbar \omega_{\mathbf{q},q_z})\theta(k-k_F),
\end{eqnarray*}
where $V^{qm}_{\mathbf{q},q_z}$ is the quasi-mode coupling matrix
element, $\omega_{\mathbf{q},q_z}$ the extracavity photon frequency and $\alpha_{\zeta}(\mathbf{q})=\braket{A,\mathbf{q}}{N+1,\mathbf{k}, 1,\zeta}$ as defined in Eq.  (\ref{general_vector}).
Having calculated the tunneling injection rate and the radiative
lifetime for the different states, we are able to evaluate the
electroluminescence spectra. It
is convenient to introduce the normalized photon emission
distribution corresponding to each eigenstate $\ket{N+1,\mathbf{k}, 1,\zeta}$, namely
\begin{eqnarray} \label{photonic_spectrum}
L(\mathbf{q},\zeta)&=& \mathcal{N}\sum_{q_z} \lvert
\alpha_{\zeta}(\mathbf{q}) \lvert ^2 \lvert  V^{qm} _{\mathbf{q},q_z}\lvert^2
\delta(\hbar\omega_{\zeta}-\hbar\omega_{\mathbf{q},q_z}),
\end{eqnarray}
where the normalization $\mathcal{N}$ is fixed by imposing
$\sum_{\mathbf{q}}L(\mathbf{q},\zeta)=1$.Ò
 The number of photons with in-plane wave-vector $\mathbf{q}$ and frequency $\omega$
 emitted per unit time is
\begin{widetext}
\begin{equation}
\label{Nph} N_{ph}(\mathbf{q}, \omega)=\frac{1}{\pi}\sum_{\mathbf{k},\zeta}
\Gamma_{inj}(\mathbf{k},\zeta) L(\mathbf{q},\zeta)
\frac{1/\tau_{r,\mathbf{k}, \zeta}}{(\omega-\omega_{\zeta})^2 +
(1/\tau_{r,\mathbf{k}, \zeta}+1/\tau_{nr,\mathbf{k}, \zeta})^2 }~,
\end{equation}
\end{widetext}
 where the last factor accounts for the Lorentzian
broadening due to radiative and non-radiative processes.
$\tau_{nr,\mathbf{k}, \zeta}$ is the non-radiative lifetime of the
electronic excitations and $\Gamma_{inj}(\mathbf{k},\zeta)$ is
given by Eq. (\ref{injector}). Fig. \ref{fig3} reports
representative electroluminescence spectra in the case of a
broadband (panel a) and narrowband (panel b,c,d) injector. In the
broadband case,  the emission is resonant at the intersubband
cavity polariton frequencies (dashed lines) and it is significant
in a wide range of in-plane wavectors\cite{EL_model}. In contrast,
in the case of narrowband electrical injector our theory shows
that the photon in-plane momentum and the energy of the cavity
polariton emission can be selected by the resonant electron
tunneling process, in agreement with what suggested by recent
experiments \cite{Luca_PRL}.

In free-space, the quantum efficiency of electroluminescent
devices based on intersubband transitions is poor ($\approx
10^{-5}$ in the mid-infrared) due to the slow radiative
recombination of long wavelength transitions. In the microcavity
case, the efficiency of the emission from an excited state
$\ket{N+1,\mathbf{k},1,\zeta}$ is given by $(1+\tau_{r,\mathbf{k},
\zeta}/\tau_{nr,\mathbf{k}, \zeta})^{-1}$. Being
$1/\tau_{nr,\mathbf{k}, \zeta}$ essentially proportional to the
matter component of the excitation and $1/\tau_{r,\mathbf{k},
\zeta}$ to its photonic fraction, we have found that it is
possible to obtain a quantum efficiency approaching unity by
selectively injecting electrons into dressed states with a high
photonic fraction. In particular, this is achievable by avoiding
injection resonant with the central peak of the electron spectral
function in Fig.\ref{fig2}, which corresponds to states with
strong overlap with the bare electron state.

In the present theory, we have not considered the role of
electronic disorder, which is known to break the in-plane
translational invariance.  However, in the limit of large vacuum
Rabi energies (i.e., significantly larger than the energy scale of
the disorder potential), the inhomogenous broadening is expected
to have a perturbative role.

Let us point out clearly that in order to achieve a high quantum
efficiency, it is necessary to have a considerably narrow spectral
width for the injector, on the order of a small fraction ($
10^{-2}$) of the intersubband transition energy
$\hbar\omega_{12}$. This is essential in order to be able to
inject electrons selectively into the superradiant states, while
avoiding both the peak associated to the {\it dark} excitations at
the bare electron energy and the states with $k<k_F$ that can not
radiatively decay.
 In the experiments in Ref.
\onlinecite{Luca_PRL}, the spectral width of the injector (a
heavily doped superlattice) is comparable to the polariton vacuum
Rabi frequency and hence such selective excitations of the
superradiant states cannot be reached. In order to have an
injector with narrower spectral width, several electronic designs
could be implemented. For example, one can grow a "filter" quantum
well between the superlattice injector and the active quantum
well: resonant electron tunneling through the intermediate quantum
well can significantly enhance the resonant character of the
excitation. Moreover, for a given injector, improved microcavity
samples with larger vacuum Rabi frequency would allow the system a
more resonant excitation of the superradiant electronic states.

\section{Conclusions}

In conclusion, we have determined in a non-perturbative way the
quasi-electron states in a microcavity-embedded two-dimensional
electron gas. Such states originate from a Fano-like coupling
between the bare electron state and the continuum of cavity
polariton excitations. We have proven that these states can be
selectively excited by resonant electron tunneling and that the
use of narrow-band injector may give rise to  efficient
intersubband polariton electroluminescence. Our theoretical work
shows that the strong coupling to the vacuum electromagnetic field
can modify significantly the fundamental electron injection
processes.

\section{Acknowledgements}

qWe thanks Iacopo Carusotto, Raffaele Colombelli, Alberto
Santagostino, Luca Sapienza, Carlo Sirtori, Yanko Todorov and
Angela Vasanelli for discussions.

\end{document}